\documentclass[article]{elsarticle}
\makeatletter
\def\ps@pprintTitle{%
	\let\@oddhead\@empty
	\let\@evenhead\@empty
	\def\@oddfoot{\centerline{\thepage}}%
	\let\@evenfoot\@oddfoot}
\makeatother

\usepackage{lineno,hyperref}
\usepackage{threeparttable}
\usepackage{float}
\usepackage{booktabs}
\usepackage{amsmath}
\usepackage{comment}
\usepackage{color}

\usepackage{amsfonts}
\usepackage{amssymb}
\usepackage{amsmath}
\usepackage{amsthm}
\usepackage{amssymb}

\usepackage{epstopdf}
\usepackage{natbib}

\modulolinenumbers[5]
\journal{Computer Physics Communications }

\begin{document}

\begin{frontmatter}

\title{A comment on the article "\textit{Ab initio} calculations of pressure-dependence of high-order elastic constants using finite deformations approach" by I. Mosyagin, A.V. Lugovskoy, O.M. Krasilnikov, Yu.Kh. Vekilov, S.I. Simak and I.A. Abrikosov [Computer Physics Communications 220 (2017) 20–30]}

\author{Marcin Ma\'zdziarz\corref{mycorrespondingauthor}}
\ead{mmazdz@ippt.pan.pl}

\address{Institute of Fundamental Technological Research Polish Academy of Sciences,	Warsaw, Poland}
\cortext[mycorrespondingauthor]{Corresponding author}

\begin{keyword}
Ab initio calculations\sep Elastic moduli\sep Pressure effects in solids and liquids\sep Finite deformations\sep Solid mechanics\sep Deformation gradient 
\end{keyword}

\end{frontmatter}

%\linenumbers

%\section{The Elsevier article class}

Recently, I. Mosyagin, A.V. Lugovskoy, O.M. Krasilnikov, Yu.Kh. Vekilov, S.I. Simak and I.A. Abrikosov \cite{MOSYAGIN201720} presented a description of a technique for \textit{ab initio} calculations of the pressure dependence of second- and third-order elastic constants. 
Unfortunately, the work contains serious and fundamental flaws in the field of finite-deformation solid mechanics.

Finite strain tensor $\eta_{ij}$ in Ref.\cite[Eq.(6)]{MOSYAGIN201720} is incorrectly defined and even wrongly rewritten from Ref.\cite[Eq.(1.5)]{WALLACE1970301}. In nonlinear continuum mechanics it is known as Green-Lagrange strain tensor and is correctly defined as: 
\begin{eqnarray}
\boldsymbol{\eta}=\frac{1}{2}(\boldsymbol{\alpha}^T\boldsymbol{\alpha}-I) \quad  or \quad {\eta_{ij}}=\frac{1}{2}(\alpha_{ki}\alpha_{kj}-\delta_{ij}),
\label{eqn:GLs}
\end{eqnarray}
see Ref.\cite[Eq.(2.67)]{holzapfel2000nonlinear}.  

The quantity $\alpha_{ij}$ called in Ref.\cite{MOSYAGIN201720} "tensor of transformation coefficients of
the system" is crucial in nonlinear continuum mechanics and is a primary measure of deformation, called the deformation gradient.
In general, tensor \boldmath{$\alpha$} is non-symmetric and has nine components, see Ref.\cite[Eq.(2.39)]{holzapfel2000nonlinear}. 
 
Deformation gradient \textbf{$\alpha$} cannot be expressed by a function of \textbf{$\eta$}:
\begin{itemize}
	\item Tensor \textbf{$\alpha$} is in general non-symmetric while \textbf{$\eta$} is symmetric, thus the right sides of this equations in Ref.\cite[Eq.(13) and Eq.(18)]{MOSYAGIN201720} are always symmetric and in general \textbf{\textit{LHS}}$\neq$\textbf{\textit{RHS}} 
	\item Let's take so-called simple shear	deformation, where the deformation gradient can be expressed as in Ref.\cite[Eq.(2.2.54)]{Ogden1984}: 
	\begin{eqnarray}
		\boldsymbol{\alpha}=\left[
		\begin{array}{ccc}
		1&\gamma&0\\
		0&1&0\\
		0&0&1
		\end{array}\right],  
		\label{eqn:SSd}
	\end{eqnarray} 
	and thus $\eta$ (\ref{eqn:GLs}) for the simple shear deformation (\ref{eqn:SSd}) will be:
	\begin{eqnarray}
		\boldsymbol{\eta}=\left[
	    \begin{array}{ccc}
		0&\frac{\gamma}{2}&0\\
		\frac{\gamma}{2}&\frac{\gamma^2}{2}&0\\
		0&0&0
	    \end{array}\right],  
		\label{eqn:GLSS}
	\end{eqnarray}  
	and now apply \textbf{\textit{RHS}} of the third order expansion defined in Ref.\cite[Eq.(13)]{MOSYAGIN201720} to our simple shear deformation(\ref{eqn:SSd}):
	\begin{eqnarray}
		\boldsymbol{RHS}=\left[
		\begin{array}{ccc}
		\frac{16-2\gamma^2+\gamma^4}{16}&\frac{8\gamma-\gamma^3+\gamma^5}{16}&0\\
		\frac{8\gamma-\gamma^3+\gamma^5}{16}&\frac{16+6\gamma^2+\gamma^6}{16}&0\\
		0&0&1
		\end{array}\right],  
		\label{eqn:13} 
	\end{eqnarray} 
	and \textbf{\textit{RHS}} of the 4th order expansion defined in Ref.\cite[Eq.(18)]{MOSYAGIN201720}:
		
	\begin{eqnarray}
		\boldsymbol{RHS}=\left[
		\begin{array}{ccc}
		\frac{128-16\gamma^2+3\gamma^4-5\gamma^6}{128}&\frac{64\gamma-8\gamma^3-2\gamma^5-5\gamma^7}{128}&0\\
		\frac{64\gamma-8\gamma^3-2\gamma^5-5\gamma^7}{128}&\frac{128+48\gamma^2-5\gamma^4-7\gamma^6-5\gamma^8}{128}&0\\
		0&0&1
		\end{array}\right],  
		\label{eqn:18} 
	\end{eqnarray} 
and we see that \textbf{\textit{RHS}} in Eq.(\ref{eqn:13}) as well in Eq.(\ref{eqn:18}) $\not\to$ Eq.(\ref{eqn:SSd}).
	\item Let's now take the deformation gradient \textbf{$\alpha$} expressed as
	\begin{eqnarray}
		\boldsymbol{\alpha}=\left[
		\begin{array}{ccc}
		\frac{1}{\sqrt{2}}&-\frac{1}{\sqrt{2}}&0\\
		\frac{1}{\sqrt{2}}&\frac{1}{\sqrt{2}}&0\\
		0&0&1
		\end{array}\right].  
		\label{eqn:RRd}
	\end{eqnarray}
	It is easy to check that  \textbf{$\eta$} (\ref{eqn:GLs}) for this deformation (\ref{eqn:RRd}) will be: 
	\begin{eqnarray}
		\boldsymbol{\eta}=\left[
		\begin{array}{ccc}
		0&0&0\\
		0&0&0\\
		0&0&0
		\end{array}\right].  
		\label{eqn:GLRR}
	\end{eqnarray} 
	and thus expansions defined in Ref.\cite[Eqs.13\&18]{MOSYAGIN201720} equal to \textbf{\textit{I}} and again $\not\to$ Eq.(\ref{eqn:SSd}). This will be the case for all deformation gradients being proper orthonormal tensors, simply a rigid rotations, see Ref.\cite[Eq.(2.2.23)]{Ogden1984}.
    We see that \textbf{$\alpha$} is not an objective strain measure since it does not vanish if the body is subjected to a rigid body motion, see Ref.\cite[Sec.(2.2)]{basar2013nonlinear}. Since a rigid rotation should not induce any stresses in a deformable body, strains should be based on deformations without being influenced by a pure rotations.
	\item Tensor \textbf{$\alpha$} has 9 components (not 6) which has been forgotten in Ref.\cite[Appendix B]{MOSYAGIN201720}, S1,$\ldots$,S11
\end{itemize}

To conclude, the authors have forgotten about the non-objectivity of the deformation gradient and its non-symmetry. In their expansions they should use symmetric and objective tensors of deformation, such as stretch tensor \textbf{U}, see Ref.\cite[Sec.(2.4)]{basar2013nonlinear}.
%See \cite{chaves2013notes}

\section*{References}

\bibliography{CPCbibfile}

\begin{thebibliography}{1}
\expandafter\ifx\csname url\endcsname\relax
  \def\url#1{\texttt{#1}}\fi
\expandafter\ifx\csname urlprefix\endcsname\relax\def\urlprefix{URL }\fi
\expandafter\ifx\csname href\endcsname\relax
  \def\href#1#2{#2} \def\path#1{#1}\fi

\bibitem{MOSYAGIN201720}
I.~Mosyagin, A.~Lugovskoy, O.~Krasilnikov, Y.~Vekilov, S.~Simak, I.~Abrikosov,
  \textit{Ab initio} calculations of pressure-dependence of high-order elastic
  constants using finite deformations approach, Computer Physics Communications
  220~(Supplement C) (2017) 20 -- 30.
\newblock \href {http://dx.doi.org/10.1016/j.cpc.2017.06.008}
  {\path{doi:10.1016/j.cpc.2017.06.008}}.

\bibitem{WALLACE1970301}
D.~C. Wallace, Thermoelastic theory of stressed crystals and higher-order
  elastic constants*, Vol.~25 of Solid State Physics, Academic Press, 1970, pp.
  301 -- 404.
\newblock \href {http://dx.doi.org/10.1016/S0081-1947(08)60010-7}
  {\path{doi:10.1016/S0081-1947(08)60010-7}}.

\bibitem{holzapfel2000nonlinear}
G.~Holzapfel, {Nonlinear Solid Mechanics: A Continuum Approach for
  Engineering}, Wiley, 2000.

\bibitem{Ogden1984}
R.~W. Ogden, {Non-{L}inear {E}lastic {D}eformations}, Dover Publications, 1984.

\bibitem{basar2013nonlinear}
Y.~Basar, D.~Weichert, {Nonlinear Continuum Mechanics of Solids: Fundamental
  Mathematical and Physical Concepts}, Springer Berlin Heidelberg, 2013.

\end{thebibliography}

\end{document}